# Book Review: *Background Independence in Classical and Quantum Gravity,* by James Read[1]

Sebastian De Haro

Institute for Logic, Language and Computation and Institute of Physics
University of Amsterdam
s.deharo@uva.nl

1 October 2024

## Abstract

This is a review of James Read's insightful book, *Background Independence in Classical and Quantum Gravity*. The book introduces various notions of background-independence which it then makes precise, and uses to make verdicts about background-independence on a wide range of examples of spacetime theories in both classical and quantum gravity. This short book is, in several ways, a worthy example of how technical philosophy of physics ought to be done. I first discuss the content, then raise a number points where I disagree with the book.

## Introduction

Background-independence has been a much-debated topic in spacetime theories. One of the main lessons of the general theory of relativity is that spacetime is not fixed, as in Newton's theory, but is dynamical. Since the shape of a spacetime depends on its matter content, the relation between geometry and matter is dynamic. Thus there is no privileged spacetime background on which physics is to be done: unlike the cases of Newtonian space and time, and special relativity's Minkowski spacetime.

Much has been made, especially in recent discussions of quantum gravity, of this rough idea of 'background-independence'. More precisely, a basic conception of background-independence is that a theory is background-independent if and only if it contains no absolute object. According to Anderson (1967: p. 83), an absolute object is a geometric object that is the same (up to isomorphism) in all of the theory's models. In special relativity, the metric that defines the Minkowski spacetime is such an absolute object. But in general relativity, the metric is not determined to have any particular form across different models, and so it is not an absolute object. In this intuitive sense, general relativity is background-independent.

---

[1] This is a longer version of a book review that will appear in *The British Journal for the Philosophy of Science*.

In quantum gravity, background-independence has sometimes been elevated to the status of a fundamental principle that future theories ought to satisfy. Furthermore, as James Read (2023: p. 132) notes, declarations of background-independence are sometimes made "from the hip", i.e. on the basis of intuitions rather than precise definitions, and without a proper analysis of the detail of the examples for which verdicts are being announced.

Read has set out to remedy this situation, by adapting from the literature six relevant notions of background-independence that he makes precise, and that he masterfully uses to make verdicts about background-independence on a wide range of examples of spacetime theories.

This short book is, in several ways, a worthy and admirable example of how technical philosophy of physics ought to be done. For it both gives precise definitions of the various notions used, and lays them against examples from both classical and quantum gravity. Thus compared to previous discussions of background-independence, Read's approach is refreshing. It is also commendable that Read engages with the current physics literature.

A good number of Read's verdicts agree with sensible views about the theories involved. For example, there is a majority verdict that Newtonian gravity theory and its geometric cousin (i.e. the Newton-Cartan theory) are background-dependent. General relativity comes out as being background-independent, as do some of its quantum versions (see Section 1). There are also some surprises: for example, there is a background-independent cousin of the Newton-Cartan theory; and Kaluza-Klein theory is not diffeomorphism invariant and yet it is background-independent. When some definitions are imprecise or give clearly incorrect verdicts in some examples, this is sometimes used to further improve the definition at issue. Thus Read builds an overall convincing case that the verdicts that are given by the majority of his definitions of background-independence are correct, and thus that at least some of the conceptions of background-independence that he discusses are valuable.

## 1. Contents of the book

Chapters 1 and 2 are brief introductions. Chapter 3 discusses the main players in the rest of the book: general covariance, diffeomorphism invariance, absolute objects and the six main notions of background-independence for classical spacetime theories. With these tools in hand, Chapter 4 goes on to study and give verdicts on background-independence in eight main examples: Newtonian gravitational theory, Newton-Cartan gravity (in two different forms), general relativity, teleparallel gravity, Kaluza-Klein theory, and two Machian relationist theories (scale-invariant particle dynamics and shape dynamics). Read summarises his findings in a helpful table (p. 93).

Chapter 5 discusses quantum gravity: perturbative string theory, AdS-CFT duality, and loop quantum gravity. Here again, the verdicts are sensible, with some of them going against often-repeated claims in the literature. Thus AdS-CFT (anti-de Sitter-conformal field theory) duality comes out as background-independent, in a very robust sense that contradicts some of the "from the hip" declarations in the literature. The verdict on perturbative string theory is not unexpected: it comes out as background-dependent, but only on the assumption that the spacetime metric is flat or close to flat. By contrast, a more general perturbative formulation of string theory on a curved spacetime would be background-independent.[2] Finally, loop quantum gravity comes out as background-independent, although Read argues that some of the literature is incorrect in claiming that it is 'manifestly' so (p. 129). Chapter 6 contains Read's conclusions about what the book has achieved and about the value of his analysis.

Overall, Read makes a convincing case that a number of the definitions of background-independence that he has adapted from the literature give correct verdicts: and that this topic can, and should, be addressed with precision. The book sets not only the agenda, but also the standard required, for discussions of background-independence in the years to come.

But there are two broad areas where I find the book to be less convincing. The first concerns a number of specific technical points (Section 2); the second concerns some aspects of the overall methodology and stated aims of the book (Section 3).

## 2. Three specific points

About the specific details, there are several points where there is room for disagreement. I will here mention three such. The first concerns a particular claim about general relativity that Read endorses. The second is about theoretical equivalence, a topic whose discussion in the book is, in my view, under-developed. The third is about Read's critique of other authors, including the present one.

(1) Read endorses Pitts's (2006) claim that general relativity has an absolute object, namely the determinant of the metric. The reason for this claim is that, for any model of the theory, and for an adequate choice of coordinates, the value of the determinant of the metric, in the neighbourhood of any point, can be set to $-1$ (alternatively, in the language of active transformations, this can be done by using an appropriate diffeomorphism of the manifold).

But, as I explain below, I do not think that this implies that there is an absolute object in general relativity. Think for example of the Minkowski metric, which has the same value in every model of special relativity.[3] The case of the determinant of the metric in general relativity is different: here, the correct statement is that, for a particular choice of coordinates, the value of the determinant can *locally* (i.e. in the neighbourhood of a point) be set to $-1$. There are problems with this, because as I argue below, being able to reach a particular value by a specific diffeomorphism is not a sufficient condition for a quantity to be an absolute object.

Let me spell out this critique in more detail. It is by no means meant as definitive. (Readers who are not interested in these details may move ahead to point (2).) Thus I consider these three important points to be not settled: nor have they, to the best of my knowledge, been discussed in any detail in the literature. But having said this, it seems that the three points combined *do* make a strong case that the determinant of the metric is not an absolute object. As follows:

(i) The determinant of the metric can be set to $-1$ *only in the neighbourhood of a point* (Pitts, 2006: p. 366). For, in general, there is no choice of coordinates that can make it have the value $-1$ everywhere. Thus the putative absolute object *is not globally defined*; its existence depends on a particular choice of coordinates on a particular region of spacetime. Furthermore, these regions differ between models. And so, it can hardly be called an 'absolute object'.

(ii) There is nothing special about the value $-1$ of the determinant: by a choice of coordinates, it can be set to *any negative value* $-x$, where $x$ is any positive number.[4] Thus there is no meaningful sense in which having this value makes the determinant into an absolute object, since it can have whatever negative value one pleases. And there is no a priori reason that says that one should choose coordinates such that the determinant has the value $-1$. There is nothing absolute about the choice, and nothing absolute about having that particular value, or any other value.

(iii) The value $-1$ has significance only insofar as it reflects the value of the determinant of the *metric in the tangent space* defined at each point, that the metric of general relativity approaches locally in Riemann normal coordinates. But it is obscure why the tangent space should set the special value.[5]

---

[2] There is an important difficulty that Read does not mention (although he does point to other difficulties, such as how to move from the excited states of a quantized string to a coherent state of such strings). Namely, it is an open problem how to give a supersymmetric version of perturbative string theory in an arbitrary spacetime. (For a discussion of some of the difficulties, see Schwarz (2020: p. 2), which discusses the Green–Schwarz string formalism in an AdS spacetime, and thereby illustrates how supersymmetric string actions are currently being constructed on a spacetime-by-spacetime basis.) Since supersymmetry is widely believed to be a requirement for a well-defined perturbative string theory, and since such a theory in an arbitrary spacetime is currently lacking, the verdict that perturbative string theory on a curved spacetime is background-independent (Read, 2023: p. 110) may strike one as somewhat premature: since this will depend on the details of that theory.
[3] In Read's (2023: p. 22) terminology, it is an absolute field, and special relativity is background-dependent on Read's Definition 9, as well as on the other definitions.

Anderson's (1967: p. 83) absolute objects are defined *globally*, and his definition does *not* characterise the determinant of the metric as an absolute object: hence my point (i). An alternative definition by Friedman (1983: p. 60) is local, and implies that the determinant *is* an absolute object (Pitts, 2006: p. 366; Sus 2008, p. 72).[6] But if so, then two conclusions follow. First, it follows that this issue about the determinant is not a good reason to view Anderson's definition as deficient, since the determinant is not an absolute object à la Anderson: which undercuts an important motivation to search for alternative definitions. Thus it would seem that correcting this verdict, and distinguishing clearly between Anderson's and Friedman's definitions, leads to different verdicts for Definition 5 in Read's (2023) table on p. 93. Second, Friedman's local definition of an absolute object is *too weak*, since it qualifies virtually every theory as background-dependent. Thus I argue that, if Pitts' interpretation of Friedman's definition is correct, then Friedman's definition is an error.[7] This leaves us with Anderson's definition, and with the truth of (i).

---

[4] This is because, under a change of coordinates (alternatively, under a diffeomorphism of the manifold), the determinant of the metric transforms by a single conformal factor. Thus, by a judicious choice of conformal factor, one can set the determinant to have *any negative value* $-x$, for $x > 0$ (but, as (i) emphasizes, this can only be done locally).

[5] It follows that, if one wishes to speak of an absolute object in general relativity, a reasonable candidate would be the tangent space at each point. However, since the tangent space is a mathematical construct rather than a physical space, it does not deserve the qualification of 'background' that is used in discussions of background-independence. De Haro and De Regt (2018: p. 658) call this an 'auxiliary space' (there is of course a physically significant principle that is associated with it: namely, the strong equivalence principle). For example, one would also not say that the set of real numbers that is used in general relativity, with a notion of a distance on them, is a fixed background. Here, the consideration of interpretative issues, for which I plead in Section 3, is relevant.

[6] I thank James Read and Adan Sus for a discussion of these points.

[7] As Pitts (2006: p. 350) discusses, Friedman announces his definition of an absolute object as a "clarification" of Anderson's definition, but in doing so he introduces crucial changes with little comment.

I have mostly used passive language, talking about choices of coordinates and I only gesturing at the active interpretation in terms of diffeomorphisms. But so far as I can tell, interpreting the diffeomorphisms required to set the determinant to $-x$ as active diffeomorphisms (one diffeomorphism for each region, and also very specific *sets of diffeomorphisms* for different models) does not change these three arguments nor justify the claim that the determinant of the metric is an absolute object.

(2) The theme of theoretical equivalence, that the book engages with at various points, seems to be under-developed. First, the book's Definition 6 of background-independence relies on a conception of theoretical equivalence (given in Definition 7): a theory is background-independent iff it has no formulation that features fixed fields. This then implies that, in assessing whether a theory is background-independent, one needs to verify that it is so for each and every one of its formulations. And this does not seem feasible in practice. Thus a related methodological consideration (namely, that one wishes to assess the background-independence of a theory formulation on its own terms, irrespective of other theory formulations) prompts Read to drop this dependence on the equivalent reformulations in his later Definitions of background-independence.

But it being a property of a particular *theory formulation* seems to go against the spirit of background-independence as a property of *theories*. And so, it seems like a weak reason for dropping the relevance of equivalent formulations in the conception of background-independence.

Furthermore, it is not clear that Read's negative formulation of background-independence in Definition 6 (i.e. 'a theory is background-independent iff it has no formulation which features fixed fields') was correct to begin with. For, if background-independence is a property of a whole theory, rather than of an individual theory formulation, then it should suffice to find *one* formulation in which the theory is background-independent, for the whole theory to be background-independent.[8] In other words, being manifestly background-independent in one formulation should secure that a theory is background-independent in all other theoretically equivalent formulations, even if this is not immediately apparent for each individual one. After all, theoretically equivalent formulations of a theory are mere reformulations; they are 'the same theory, written in different words': and a *substantive physical notion like background-independence* should not depend on the words used to formulate a theory.

Think of the analogy with the gauge-invariance or coordinate-invariance of a given property in a gauge- invariant or coordinate-invariant theory: to demonstrate that some quantity has a gauge-invariant or a coordinate-invariant property, one does not need to demonstrate this property for every possible gauge or coordinate system in which the quantity can be expressed. Rather, it suffices to find, among all those equivalent formulations, one formulation where the property is instantiated.

---

[8] Because Definition 6 does not distinguish between physical and auxiliary fields (or their formal correlates, namely common core and specific structure), which is an essential distinction when discussing (even just formally) theoretical equivalence, the case in the opposite direction is not true. In other words, if the theory *looks* background-dependent in one formulation (in the sense that it features a fixed field), it need not feature fixed fields in all other formulations, and the theory need not be background-dependent. Thus this apparent background-dependence (i.e. a formulation's featuring a fixed field) is a property of a particular formulation: if, in formulating one's theory, one adds additional structure that is not physically significant but is nevertheless fixed, then that formulation can still be theoretically equivalent to the original one, and the theory need not be rendered background-dependent. Thus I argue that it should be sufficient to find one theoretically equivalent formulation in which the theory does not feature fixed fields, for the whole theory to be background-independent. See point (A) in Section 3 about the need to consider explicitly matters of interpretation when discussing background-independence.

Thus, on any account, background-independence must be compatible with theoretical equivalence. But this is not to claim that I have a worked-out view of the compatibility between background-independence and theoretical equivalence (hence my earlier use of the word 'should'). Rather, I am saying that Read's dismissal of the question of theoretical equivalence without further discussion, based only on the methodological principle stated above, seems too quick.

(3) In a section titled *Coda*, Read (2023: p. 123) correctly criticizes a footnote in a paper I wrote a decade ago (mea culpa!) for conflating the uses of the phrases 'general covariance' and 'diffeomorphism invariance' (De Haro, 2017: p. 115). And on several central points, Read makes an effort to correctly represent work by other authors. But on various other points, his reading of my work seems a bit off and unrecognizable. For example, in his discussion of my minimalist criterion of background independence, he misconstrues the reason for one of the conditions for such independence, which he then drops.[9] And on p. 120, he claims that I am committed to a notion of a theory where the boundary conditions are arbitrary. I mention these corrections, not just in "self-defence", but also because, since the book criticizes other authors, the above leaves me wondering about the degree to which its critical engagement with other authors' work is also accurate.

## 3. Methodology and aims

Among the more general issues, I have two methodological ones and some further comments about the stated aims of the book.

(A) The first methodological issue is that the book's treatment of background-independence is mostly formal, and engages little with interpretation.[10] This is surprising, especially given that, in his other work on symmetries and dualities, Read stresses the importance of the interpretative aspects of physical theories, in addition to the formal ones (see e.g. Pooley and Read (2021), Read (2016), and Read and Møller-Nielsen (2020)). I here give two instances of this methodological issue:

One of the book's main conceptions of background-independence (Definition 10, p. 24) requires that the solution space is determined by a generally covariant action, all of whose dependent variables are subject to Hamilton's principle. Since, up to that point, definitions of background-independence had made no reference to Hamilton's principle, one would expect some motivating discussion as to why a variational principle is required here (the more so, because Read correctly insists on the fact that not all theories in physics have an associated action from which the field equations can be derived). While one can easily think of reasons of *convenience* for introducing the variational principle (e.g. allowing an extension of a classical theory into a quantum field theory by using Feynman's path integral over trajectories), there is no clear principle that *requires* a treatment in terms of an action and its variation. Agreed, Read was here following Pooley's lead for the sake of being comprehensive, while acknowledging the limits on the scope of any definition in terms of variational principles. But a discussion of the physical motivation for focussing on a variational principle would have been desirable.

---

[9] Read (2023: p. 118) suggests that the reason why I add condition (b) to my conception of minimalist background-independence is because I wish to recover general relativity in the classical limit. But instead, this condition is required because it is not sufficient to consider the background-independence of the field equations. One must also consider the background-independence of the *quantities*, which is done by condition (b). Read first reformulates my condition (b), thereby dropping the most important part of it (namely, the mention of the lack of a background metric). He then drops condition (b) altogether. In the conception of a theory as a triple of states, quantities and dynamics that I used in those papers (De Haro et al., 2016; De Haro, 2017), dropping this condition would surely be incorrect, because a theory is not determined by the field equations alone, but also by the quantities that are defined on the state space (and for spacetime theories, the question of how to define a complete set of quantities or observables is a very non-trivial one—that Read's book does not mention).

[10] As I mention below, Read does stress that some of the definitions that he adapts from the literature have an interpretative aspect. But, so far as I can tell, he does not really make these interpretative aspects bear on his own theorising about background-independence.

A more central point is that Read does not say whether we should take background-independence to be a formal i.e. non-interpretative notion, or whether it is also interpretative. (Think of the analogy with symmetry: various formal methods can give us an account of the notion of a symmetry, but only an interpretative analysis can tell us whether a given symmetry is 'gauge'.) Read declares in the Introduction: 'I am concerned with the *formal* features' of background-independence (p. 2). But since (as he also stresses) some of his definitions include an interpretative aspect, this seems to be a methodological choice, rather than the expression of a firm conviction that background-independence must be formal. Thus I find his silence about this central question disconcerting. A book on background-independence merited a more substantial discussion of this crucial point.

My own view is that background-independence *does* have an interpretative aspect. And Read's analysis of the examples suggests that a two-step procedure can be successful. Here, I mean 'two-step procedure' by analogy to other notions in philosophy of science, such as for example theoretical equivalence. The first step in giving a conception of theoretical equivalence is a suitable formal notion of equivalence of theories (for example, duality, or categorical equivalence) with only minimal constraints on interpretation. The second step is an interpretative criterion of equivalence (for example, that, in terms of a referential semantics, the two theories have the same domains of application, with formally equivalent terms referring to the same items in the domain of application). While there is debate about both types of criteria, distinguishing these two steps facilitates the analysis, since it allows focussing on the relevant discussions at each step.

Since Read's formal definitions of background-independence are successful in terms of scope and reasonableness of his verdicts (especially his Definitions 9 and 10), I take this as an indication that the two-step approach to background-independence is the preferred one.

(B) The second methodological issue concerns the number of alternative definitions of the same concept, especially 'background-independence', that the reader is asked to keep track of. We are offered little guidance as to whether a new definition that is introduced is better than the previous one. (As I discussed above, this is for example the case with the introduction of Definition 10, which incorporates the unmotivated idea of a variational principle.) In his Conclusion, Read celebrates this 'pluralism': 'there is a plurality of definitions, each of which is helpful in enriching our understanding of the workings of our physical theories: and it is this enrichment, rather than identifying any 'true' notion of background-independence, which is the goal' (p. 133).

This type of pluralism is fine, so far as it goes. But not searching for the one true notion of background-independence in one thing; and carrying along six classical notions (plus four quantum ones), without providing guidance about their relative merits and their acting as motivations for introducing new principles, is another.

For example, although the various verdicts are illustrated for different theories, some of the definitions clearly give incorrect verdicts (this is the case with Definition 5, which classifies virtually any spacetime theory as background-dependent).[11] Read continues to carry Definition 5 along, rather than dropping it as inadequate. And, despite the 'majority verdicts' that Read emphasises, the various notions have minority disagreements about several central examples that were worth trying to explain. Surely more guidance and judgment about which notions to keep, and about how the different notions rank relative to each other (perhaps for different types of theories), would have been desirable.

This leads in to a more general point about the stated aims of the book. The Introduction advertises that background-independence is a heuristic for theory construction, rather than a norm to be fulfilled (p. 2). And the Conclusion reassures us that background-independence is a guiding principle for the construction of new physical theories, and that 'the more nebulous 'spirit' of background-independence, loosely construed, can also be used as such a guiding principle' (p. 134). But notwithstanding this assurance, one is left wondering whether background-independence thus construed can realistically be considered a guiding principle for theory construction: especially given the number of different classical conceptions of background-independence, the lack of principles for when to use each of them, and the very limited attention to physical interpretation. '[G]iven the flexibility and potential lack of univocity' (p. 131) of the notion of background-independence, which Read himself admits, how useful a guide to theory construction by physicists can this notion be?

As a self-declared empiricist, Read denies that background-independence should be understood as a rationalist principle or necessary condition for good theory construction (p. 134). But an empiricist can analyse and discuss the pros and cons of a given conception of background-independence without being thereby committed to *a priori* knowledge of how an ideal theory should be constructed. This is what the debates about the best conceptions of many notions used by philosophers of physics are about: symmetries, manifolds, gauge, entropy, etc. Thus I think that such work is yet to be done for background-independence, though Read's book gives a good formal basis on which to build.

To sum up: Read has given us a wonderful analysis of the formal aspects of various conceptions of background-independence, with minimal discussion of their interpretation. Thus the enterprise of discussing with comparable precision the adequacy of each of these notions (perhaps in various contexts), and the interpretative and epistemic principles according to which we choose one notion over another, is yet to begin.

---

[11] But see the subtleties discussed after point (1)-(iii) in Section 2, where I argued that the analysis of the determinant of the metric suggests that Friedman's, rather than Anderson's, definition should be dropped.

## Acknowledgements

I thank James Read and Adan Sus for discussions of the content of this review, and Jeremy Butterfield and Enrico Cinti for comments.
## References

Anderson, J. L. (1967). *Principles of Relativity Physics*. New York: Academic Press.

De Haro, S., Mayerson, D. R., and Butterfield, J. (2016). 'Conceptual aspects of gauge/gravity duality', *Foundations of Physics,* 46: pp. 1381–1425.

De Haro, S. (2017). 'Dualities and emergent gravity: Gauge/gravity duality'. *Studies in History and Philosophy of Modern Physics,* 59, pp. 109-125.

De Haro, S. and De Regt, H. W. (2018). 'Interpreting Theories without a Spacetime'. *European Journal for Philosophy of Science,* 8, pp. 631-670.

Friedman, M. (1983). *Foundations of Space-Time Theories*. Princeton University Press.

Pitts, J. B. (2006). 'Absolute objects and counterexamples: Jones–Geroch dust, Torretti constant curvature, tetrad-spinor, and scalar density'. *Studies in History and Philosophy of Modern Physics,* 37, pp. 347–371.

Pooley, O. and Read, J. A. M. (2021). 'On the Mathematics and Metaphysics of the Hole Argument'. Forthcoming in *The British Journal for the Philosophy of Science*.

Read, J. (2016). 'The Interpretation of String-Theoretic Dualities'. *Foundations of Physics,* 46, pp. 209-235.

Read, J. and Møller-Nielsen, T. (2020). 'Motivating Dualities'. *Synthese*, 197, pp. 263-291.

Sus, A. (2008). *General Relativity and the Physical Content of General Covariance*. PhD thesis, Universitat Autònoma de Barcelona.

Schwarz, J. H. (2020). 'The $AdS_5 \times S^5$ Superstring'. Proceedings of the Royal Society, A 476: 20200305, pp. 1-8.